# An Overview of Potential Medical Applications of Microwave Kinetic Inductance Detectors

Cáthal McAleer, Gary Donegan, Colm Bracken, Oisín Creaner,
Mario De Lucia, Tom Ray, Gerhard Ulbricht

***Abstract*—Microwave Kinetic Inductance Detectors (MKIDs) have been proven to be a versatile detector technology and are very promising for several applications in medical imaging. Deep imaging into human tissues often requires detectors with low background noise, high detection efficiency, broad wavelength sensitivity or single-pixel energy resolution. MKIDs offer a unique combination of valuable features for medical imaging but have so far not been discussed much for modern medical instrumentation. This represents an exciting opportunity to investigate the potentially powerful advantages MKIDs could offer for clinical studies in several areas. The most promising fields for MKIDs in medical imaging include fluorescence microscopy, near-infrared spectroscopy, energy-resolved medical x-ray imaging and synchrotron x-ray microscopy. We present the respective advantages MKIDs offer for these fields and compare them to current detector state-of-the-art.**

## I. INTRODUCTION

MICROWAVE Kinetic Inductance Detectors (MKIDs) are novel superconducting single photon detectors that have been proven powerful and highly versatile in astronomical instrumentation. They are also very promising for medical imaging for multiple use cases. Deep imaging into biological tissues is limited by scattering and absorption and therefore requires detectors with low background noise, high detection efficiency and ideally broad wavelength sensitivity. Semiconductor based photo-detectors (CCDs, CMOS, Photomultiplier Tubes (PMTs), etc.) have reached their physical limitations based on semiconductor bandgap, and often require the use of scintillators for higher energy photons, degrading spatial resolution. They are also typically limited to small wavelength ranges, and also often require high radiation doses. In comparison, MKIDs promise to offer significant advantages for modern medical instrumentation, for example $\mu s$ time resolution, lack of dark counts or single pixel and single photon energy resolution.

MKIDs exhibit many unique capabilities for medical imaging but are still in an early development phase and have so far not been discussed much for modern medical instrumentation. Their significant detector advantages compared to currently used technologies have the potential to make MKIDs promising for many challenges in medical imaging. Here we will present four exemplary cases in their respective chapters where MKIDs have the potential to provide the largest impact:

II. Fluorescence Microscopy
III. Near-Infrared Spectroscopy
IV. Energy-Resolved Medical X-Rays; and
V. Synchrotron X-Ray Microscopy

## II. FLUORESCENCE MICROSCOPY

Fluorescent microscopy is a powerful tool to understand biological processes occurring at the cellular level. The technique can be used for living (in-vivo) or encased/fixed cellular imaging and is often used in, for example, virology or embryonic development. Specimens are tagged with fluorescent markers (so called fluorophores) that bond to specific cellular structures and fluoresce when illuminated with one or more excitation wavelengths, usually in the visible to near-IR range. The fluorescence is then captured by either detector arrays or single-pixels via raster scanning (see Fig. 1).

Multiple approaches how to excite fluorophores and capture fluorescent signals exist, the most common being one-photon or two-photon excitation. One-photon fluorescent microscopy [1] gets its name from exciting fluorophores with a single

Received 10 September 2025; revised 16 February 2026; accepted 24 June 2026. Date of publication 26 June 2026; date of current version 14 August 2026. This work was supported in part by the Taighde Éireann– Research Ireland, under Grant 21/FFP-P/10213 and Grant 15/IA/2880 and in part by the Astronomy & Astrophysics Section, School of Cosmic Physics, Dublin Institute for Advanced Studies a grant. (*Corresponding author: Cáthal McAleer.*)
Cáthal McAleer is with the Department of Physics, Maynooth University, W23 F2H6 Maynooth, Ireland, and also with the School of Cosmic Physics, Dublin Institute for Advanced Studies, D04 C932 Dublin, Ireland (e-mail: cmcaleer@cp.dias.ie).
Gary Donegan, Colm Bracken, and Gerhard Ulbricht are with the Department of Physics, Maynooth University, W23 F2H6 Maynooth, Ireland.
Oisín Creaner is with the School of Physical Sciences, Dublin City University, 9 Dublin, Ireland.
Mario De Lucia is with the INFN sezione di Pisa, 56127 Pisa, Italy, and also with the Universita di Pisa, 56126 Pisa, Italy.
Tom Ray is with the School of Cosmic Physics, Dublin Institute for Advanced Studies, D04 C932 Dublin, Ireland.

wavelength in the range of 300 nm to 700 nm, and therefore just requires a single photon per fluorophore. It also requires pinhole apertures in order to avoid out-of-focus fluorescence and noise. Two-photon fluorescence microscopy [2] makes use of two simultaneously absorbed, lower energy photons to excite a single fluorophore, which allows for longer wavelengths up to 1300 nm to be used, achieving greater imaging depths and intrinsic optical sectioning (depth resolution) as the emission occurs only at the focal point, even without pinholes. In both cases, typical fluorophore emission wavelengths range from optical to near-IR (500 nm up to 1700 nm [3]).

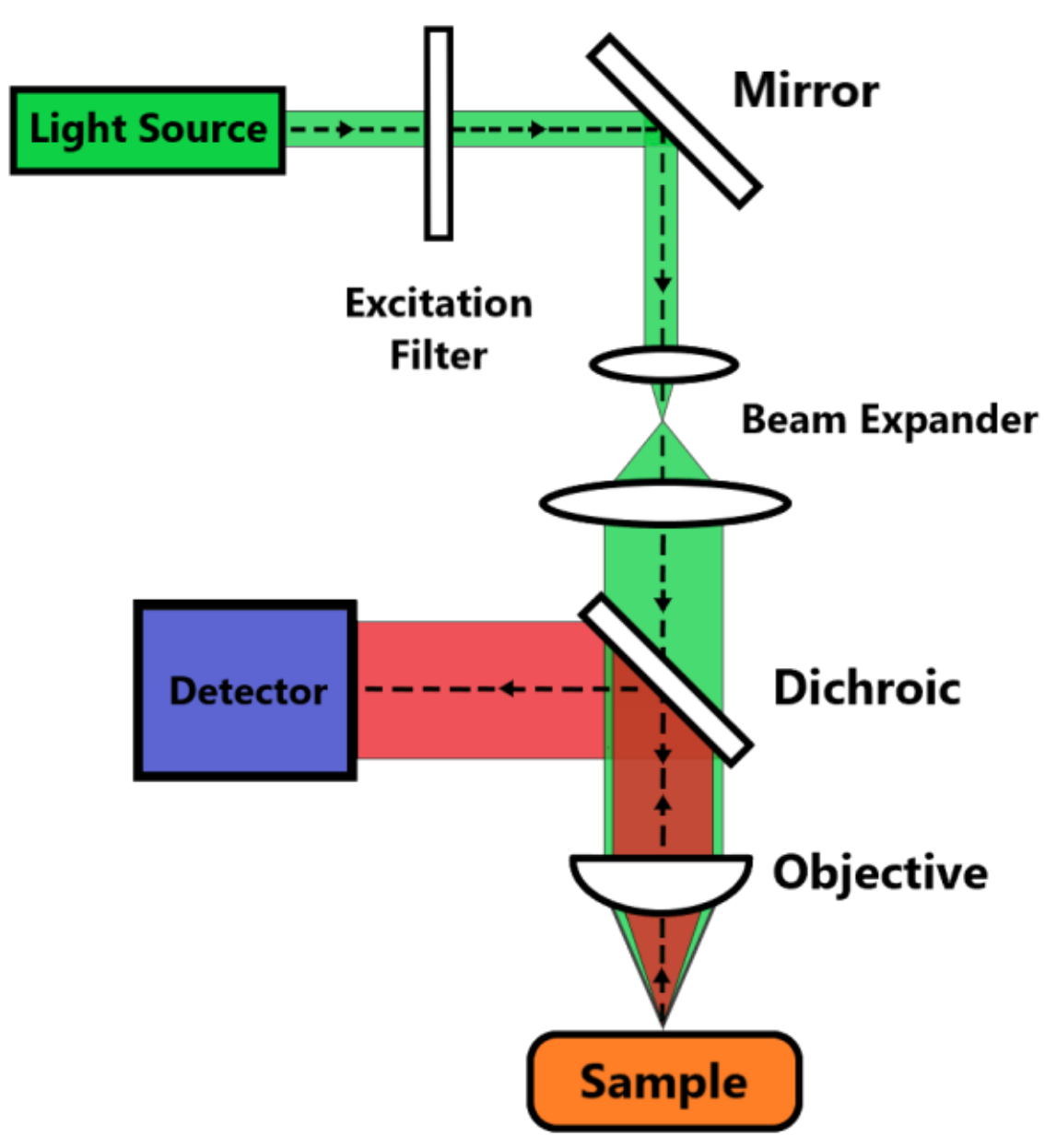

**Fig. 1:** Schematic of the fluorescence microscopy principle.

*A. State of the art*

Current fluorescence microscopy [4] mostly employs large format CCD arrays or single-photon counting PMTs if deep tissue penetration is required. CCDs suffer from background noise and therefore require higher photon intensities, risking photo bleaching and tissue damage. In areas of low-intensity fluorescence, PMTs are generally required but are bulky and suffer from dark counts [5]. PMTs are also incapable of determining the wavelength of detected photons and therefore cannot be used with multiple, simultaneous fluorophores.

*B. Advantages of MKIDS / Superconducting Photon Detectors*

PMTs struggle to achieve more than 1.0 mm of tissue penetration (see Fig. 2). Superconducting detectors, in this example Superconducting Nanowire Single Photon Detectors (SNSPDs) have already achieved similar and/or better results, while offering better noise performance. However, SNSPDs lack wavelength resolution and are therefore limited to single fluorophore observations. MKIDs on the other hand also offer single-photon counting but have virtually no dark counts at all [6], and should therefore be capable of achieving similar penetration depths. In addition, MKIDs offer single-photon wavelength resolution (of currently up to R ~ 52 [7]) and simple frequency domain multiplexing (FDM), offering a feasible way towards large detector arrays and therefore improved imaging for fluorescence microscopy. Their wavelength resolution does not increase penetration depth but allows MKIDs to simultaneously use different fluorophores and longer excitation wavelengths for both one-photon and two-photon fluorescence microscopy. MKIDs can therefore be used with different fluorophores (quantum dots, indocyanine green, … [8]), binding to different parts of living cells (so called cell organelles) and thus allow for the study of multiple sub-cellular interactions between cell organelles simultaneously in real time.

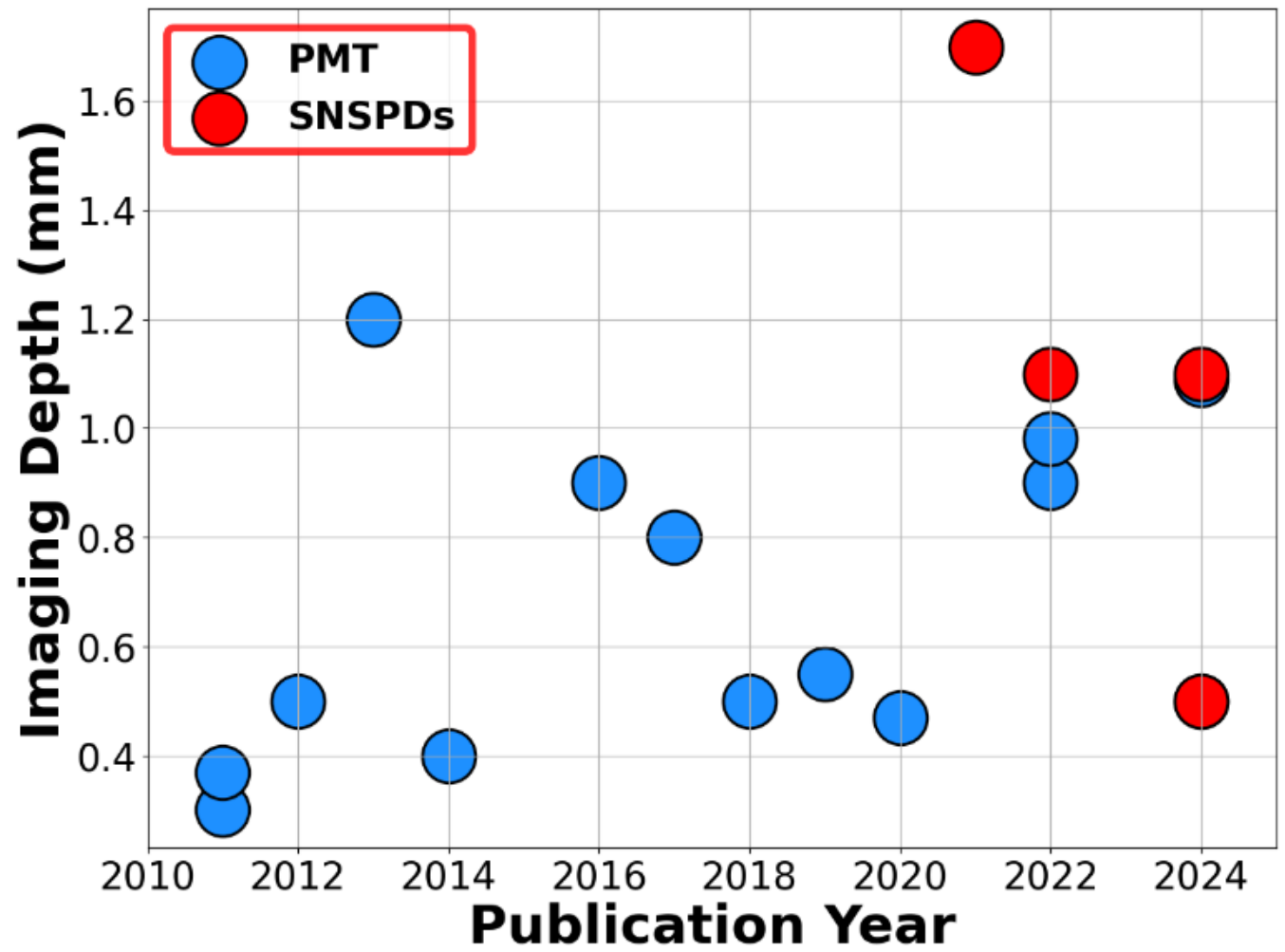

**Fig. 2:** Limits to imaging depths in fluorescence microscopy (one, two and multi-photon fluorescence), comparing PMTs and SNSPDs [9], [10], [11], [12], [13], [14], [15], [16], [17], [18], [19], [20], [21], [22], [23], [24], [25]. Excitation and emission wavelength ranged from 760 nm – 1700 nm and 515 nm – 1700 nm respectively. Due to their single photon sensitivity, MKIDs are expected to achieve similar penetration depths to SNSPDs.

## III. Near-Infrared Spectroscopy

Medical imaging utilizing Near-Infrared Spectroscopy (NIRS) is regularly used for the spectroscopic analysis of in-vivo blood-oxygen concentrations [26], which is a well-developed bio-imaging technique since the 1970's. NIRS is a non-invasive imaging technique and highly attractive for many clinical applications, for example infant brain development studies, stroke rehabilitation, psychiatry or early cancer diagnoses [27], [28], [29], [30].

Between 700 - 900 nm, there exists an optical window where near-IR light can diffuse through tissue and even bone but is absorbed significantly more by blood chromophores (see figure 3). The two main NIR chromophore absorbers in the blood are oxygenated and de-oxygenated haemoglobin [31]. Especially in brain imaging, measuring blood oxygen concentration can be invaluable to diagnose many medical conditions like hypoxia or strokes.

In brain imaging, light is either absorbed or scattered (so called Mie-scattering, as the distance between brain tissue cells is comparable to the wavelength). Light that is scattered approximately travels in a curved "banana" shape that can be detected leaving the scalp. By measuring light absorption through the tissue with at least two different NIR wavelengths between 700 and 900 nm, the relative changes in oxygenated to de-oxygenated haemoglobin concentration can be determined by solving the modified Beer-Lambert Law [32], [33], [34] as follows:

$$\log_{10}\left(\frac{I_{before}}{I_{after}}\right) = \alpha(\lambda)^{Hb}\Delta c^{Hb}D + \alpha(\lambda)^{HbO_2}\Delta c^{HbO_2}D \quad (1)$$

Here, $I_{before}$ and $I_{after}$ are the measured light intensities at the start and end of an experimental task (e.g., [35]). $\alpha(\lambda)$ is the wavelength dependent absorption coefficient of oxygenated or de-oxygenated haemoglobin and $\Delta c$ is the change in oxygenated or de-oxygenated haemoglobin concentration between the two measurements. D represents the path length between light source and detector (on the skull) and includes estimated values based on the random walk of NIR photons inside the tissue. The path length $D$ is assumed to remain constant throughout both measurements. Two measurements near both edges of the optical window allow to determine $\Delta c^{Hb}$ and $\Delta c^{HbO}$ independently.

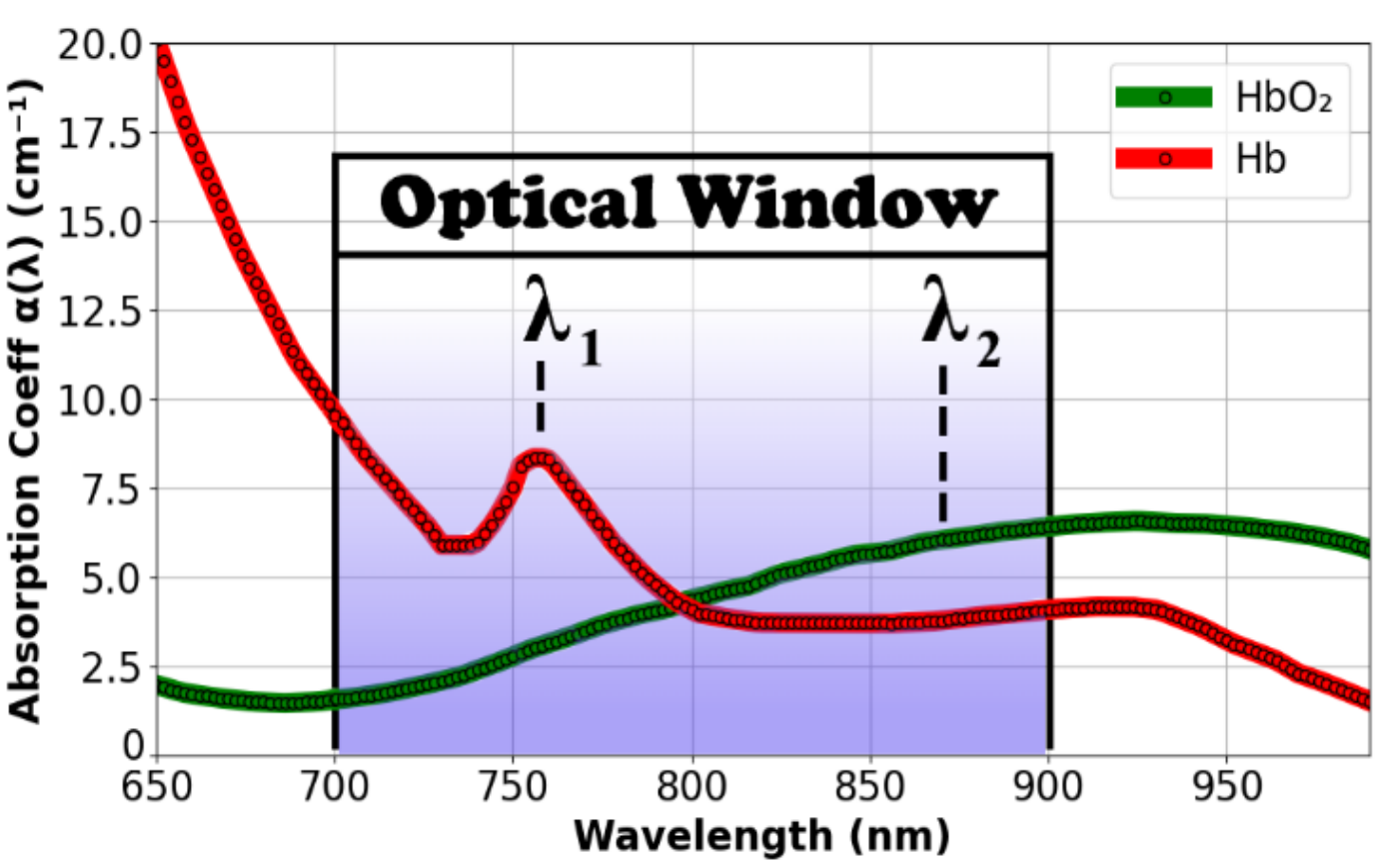

**Fig. 3.** Exponential absorption coefficient for oxygenated ($HbO_2$) & de-oxygenated haemoglobin (Hb), data obtained from [36].

Changes in haemoglobin concentrations during tasks asked of the patient can then be linked to other processes occurring in the body such as relating oxygenated haemoglobin concentration to localized neural activity in the brain (neurovascular coupling) or early tumor formation (angiogenesis) [26], [37]. As such, there is currently high interest in research in areas such as functional NIRS to monitor brain activity [38] and Diffuse Optical Tomography-NIRS used in breast cancer screenings [39].

### *A. Currently used Detectors and Limitations of NIRS*

NIRS is primarily employed in wearables where detector-source pairs (so called optodes) are placed on the head with adhesives or in wearable caps, and the patient is monitored as they perform tasks or activities. In most cases, the detectors currently used are general purpose photodiodes, while in more complex systems (such as Time-Domain NIRS [40]), avalanche photodiodes (APDs) or PMTs are employed. NIRS is a diffusive imaging technique and is therefore limited by the non-linear scattering and exponential intensity decay of photons propagating through tissues. The achievable imaging depth in most cases is roughly given by the source-detector distance (SDD) [26] as follows:

$$Imaging\ depth \approx \frac{SDD}{2} \quad (2)$$

When using NIRS with photodiodes, the main limitations come from their lower sensitivity and higher noise floor. This constrains the imaging depth (to about 1cm) while also restricting the possible number of light source / detector pairs due to larger source-detector distances required to reach this depth (see Fig. 4). As such, only superficial cortical activity may be imaged at low resolutions (ca. 1 $cm^2$ image pixel size). At these shallow depths, signal and movement artifacts such as blood flow in the scalp also become difficult to remove in post-processing and can corrupt cortical signals [41].

APDs and PMTs out-perform general purpose photodiodes by offering single-photon counting while achieving higher signal-to-noise-ratios (SNR), allowing for deeper penetration depths (3 – 4 cm) [42]. Ultra-High Density NIRS (UHD-NIRS) arrays of around 126 APD pixels have also been shown to improve spatial resolution (to about 5 $mm^2$ image pixel size) and artifact removal [43]. However, both APDs and PMTs require bandpass filters or fast-switching laser modules in order to discriminate between wavelengths required to solve for chromophore concentrations. This can prevent fast cortical changes during experimental task by the patient from monitored simultaneously. They also suffer from significant dark counts at larger source-detector distances.

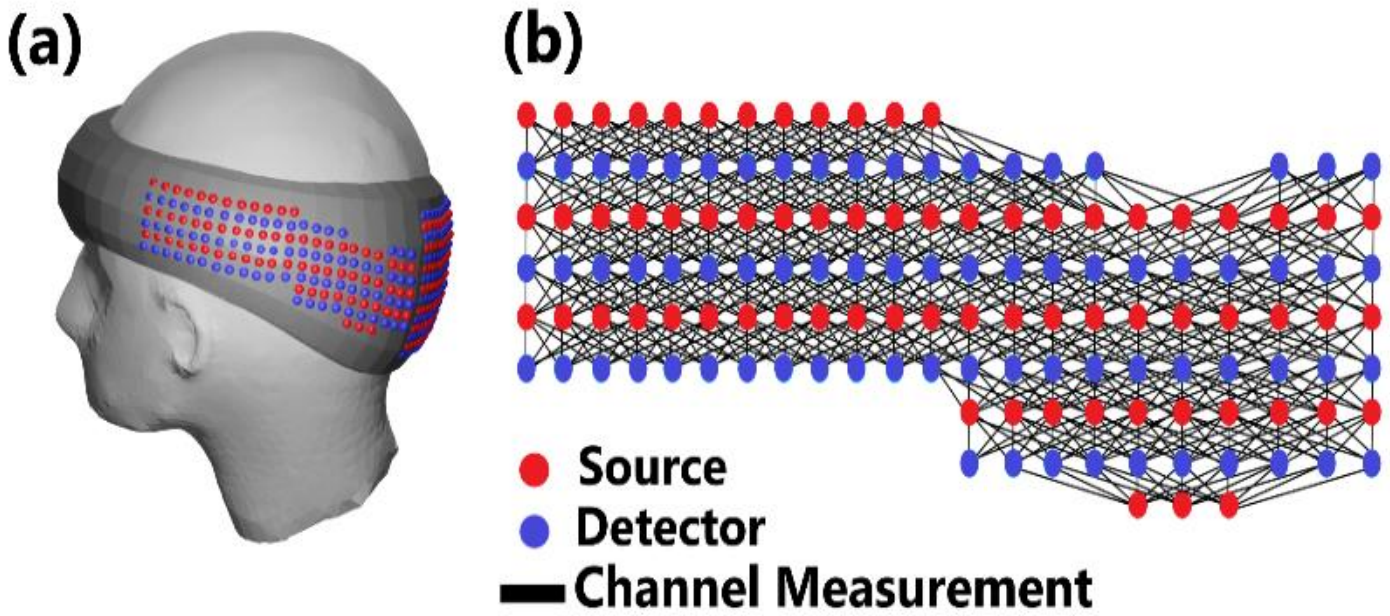

**Fig. 4.** (a) Example of a UHD-NIRS detector arrangement, based on [44]. (b) Spatial distribution of source-detector pairs across the head. Each source can be used by multiple detectors (channels), achieving different imaging depths across the head.

### *B. Advantages of MKIDs for Brain Imaging*

Brain imaging is the most established application of NIRS but is reaching limitations. Therefore, it is open to new ideas like superconducting detectors such as MKIDs. The capability of NIRS to correctly identify cortical haemodynamics (known as 'anatomical specificity') is based on the number of source-detector pairs and their spatial arrangement across the head [45]. The surface area of an adult scalp is ~700 cm$^2$ and with an achieved minimum source-detector spacing of 6.5 mm [43], approximately 2,000 detectors would be required to perform simultaneous UHD-NIRS measurements globally across the head. NIRS is also sensitive to blood dynamics occurring in the skin, contaminating signals from the brain. Small source-detector distances (< 8 mm) and higher pixel numbers can effectively sample these surface signals and remove them from the cortical signals in post-processing [46].

Competing superconducting detectors have difficulties reaching more than 500 pixels, but MKIDs can be easily multiplexed to achieve 2,000 pixels on a single feedline while providing single photon counting sensitivity with virtually no dark counts. To achieve this, MKIDs need to be operated at cryogenic temperatures and therefore require bulky cryogenic equipment. Cryostats are already widespread in modern clinical environments [47], and while the cryogenic requirements increase cost and maintenance efforts, the significant advantages offered with MKIDs effectively offset these costs. As MKIDs operate at cryogenic temperatures and cannot be placed directly on the scalp, they require optical fibres to couple between the cryostat and the patient's head. Photo-detectors are routinely coupled in this manner for NIRS applications, with fitted optical fibre caps/headgear widely used [48]. Every point across the head can therefore be easily sampled simultaneously with MKIDs, achieving a higher anatomical specificity of global cortex haemodynamics. This would allow for improvements in diagnosing and monitoring disorders and traumas that effect the entire brain, such as hypoxia, dementia and neonatal disorders. Furthermore, the large pixel numbers also allow for the use of smaller source-detector distances, enabling systematic identification and removal of scalp signal contaminants.

MKIDs also offer single-pixel wavelength resolution and even with lower resolving powers (e.g. R ~ 10), MKIDs can resolve the multiple wavelengths required to differentiate between oxy and de-oxy haemoglobin and perform basic NIRS operation. As the amount of photons to be detected in NIRS is small, this intrinsic energy resolution allows NIRS to be significantly faster compared to non-resolving detectors. Higher R could even enable to resolve other chromophores that exist in the blood such as cytochrome-c-oxidase (spectral window 830-850 nm), which would further improve diagnostic capabilities. Unlike haemoglobin, cytochrome-c-oxidase is a metabolic enzyme responsible for more than 95% of cellular oxygen uptake but requires deep NIRS penetration depths beyond ~1.5 cm to determine concentrations [49]. Resolving concentrations of this enzyme infers information on neuronal energy consumption and tissue oxygen uptake, critical in areas such as early detection of hypoxia (lack of oxygen). The challenge with cytochrome-c-oxidase lies in requiring up to 8 different wavelengths for NIRS to reliably resolve localized concentrations [50].

There are currently a large amount of detector technologies employed in NIRS and we do not intended to present the advantages and disadvantages of every detector as this would require significantly deeper discussions. The crucial advantages of MKIDs for NIRS compared to all competitors are their capability to detect single photons, their intrinsic wavelength resolution, the fact that MKIDs have virtually no dark counts and their viability to reach up to megapixel array sizes, which is unique for superconducting detectors. This unique set of capabilities is currently not shared by any other competing detector technology.

### *C. Advantages of MKIDs for NIRS for Breast Cancer Screenings*

Breast cancer is the 2$^{nd}$ most common cancer amongst women and the risk of developing breast cancer increases with age. If a tumor is screened and treated at an early stage, the prognosis of the patient will be significantly improved. X-ray mammography is currently the cornerstone of breast cancer screening, however the use of ionizing radiation can cause new lesions to form or aggravate existing tumors [51].

Diffuse Optical Tomography NIRS (DOT-NIRS) [52] has been developed as a non-invasive screening method for breast cancer. In DOT-NIRS, multispectral images are generated to characterize haemoglobin, water and lipid concentrations and used to create volumetric 3D models of internal breast tissue [37]. A larger optical window of 650 nm – 1000 nm is used as the identification of water and lipid concentration becomes necessary for cancer detection. Haemoglobin concentrations differ between malignant and benign tumours due to increased blood vessel formation and therefore can be used as markers for cancerous tissue [53]. DOT-NIRS however often suffers from a dominance of superficial surface signal contributions, particularly from a lack of measurement points. The most prevalent source of noise in DOT-NIRS arises from near-IR light scattering near the surface of the breast, contaminating deeper signals [54].

In the case of DOT-NIRS, the kilo-pixel count offered by MKIDs would significantly improve spatial resolution and could allow for better tomographic representation of internal breast tissue, which is crucial to localize even small tumour formations. It should also be noted, while DOT-NIRS is fundamentally limited to ~5 mm$^2$ spatial resolution by the mean free path of approximately 1 cm of near-IR photons in living tissue [55], the potentially large pixel number and higher SNR achieved with MKID arrays would allow for more over-sampling, improving tumour localization beyond the 5 mm$^2$ limit [56].

## IV. Energy-resolved medical x-rays

Medical x-ray machines are widely used clinically as a bio-imaging technology to screen large, internal areas of the body. Typically in single-pass instruments, x-rays in the keV range (roughly >16 keV for single pass and 60-70 keV for CT x-rays [57]) are generated via x-ray tubes (tungsten filament in a vacuum tube). Radiation is absorbed or scattered at different rates depending on tissue and bone composition and density. Monochromatic 2D images of bones and internal organs can then be generated by measuring the attenuation of incident x-rays (known as 'projectional radiography'). Transmitted radiation is detected either by scintillation detectors or luminescent storage phosphor plates read out with a scanning laser [58]. The resulting images show lower intensities for denser structures (i.e. bones) while soft tissue such as organs absorbs less radiation and shows up dark in the traditionally negative image format. The challenging visibility of soft tissue is usually remedied with Computed Tomography (CT) by utilizing a greater amount of x-ray passes at multiple angles to resolve softer tissues and to produce high-resolution, 3D images. However, a typical chest x-ray CT is ~450 times the effective radiation dosage compared to a single-pass x-ray image (27 mSv vs. 0.06 mSv respectively [59]). In fact, estimates show that 5% of all new-cancer diagnosis in the future could be CT-related [60], which demonstrates the need for less-invasive x-ray imaging techniques.

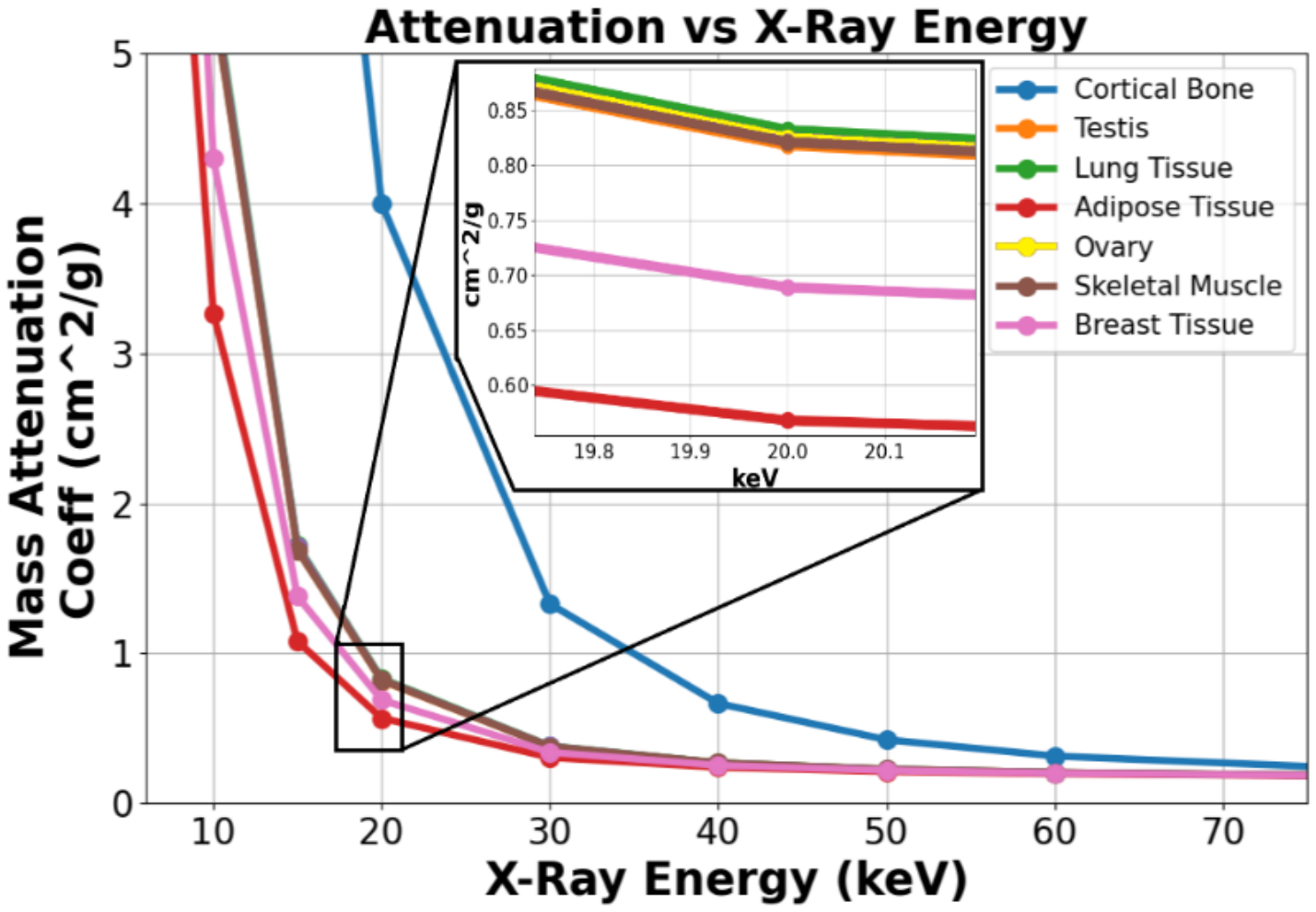

**Fig. 5.** X-ray attenuation coefficient per density for different types of human tissue. The insert demonstrates the challenge of differentiating soft body parts. Data obtained from [61].

Energy-resolved x-ray imaging (or 'multi-energy x-rays') is an attractive technique within diagnostic x-ray imaging that distinguishes soft tissues by their x-ray absorption rates to significantly increase image contrast with single-pass exposures. Organ tissue absorption rates are highly wavelength dependent (see Fig. 5) and, if irradiated with a polychromatic source, can allow for better tissue identification via spectral analysis [62]. By improving the contrast of soft tissues, energy-resolving x-ray imaging can produce images on-par with CTs while significantly lowering patient radiation dosage and improving medical diagnosis.

Energy-resolved x-ray medical imaging typically uses broadband x-ray illumination. By employing dual-layer detectors or photon counting detectors, spectral information can then be obtained in order to distinguish multiple contrast agents simultaneously. Pacella et al. [63] also suggests that for CT, multi-energy imaging could further improve morphological imaging based on intrinsic tissue contrast.

### *A. State-of-the-art in Energy-Resolved Medical X-Rays*

Energy-resolved medical x-ray imaging is still under development as an imaging technique. Early clinical trials have been performed for applications such as chest radiography to find low contrast indicators for lung cancer or digital subtraction mammography [64], [65] but significant detector limitations have been demonstrated. The challenge for the x-ray detection lies in the fact that scintillating flat panel detectors should be avoided as they degrade the spatial imaging resolution and increase detector noise [66]. Direct conversion (the absorbed x-rays producing electrical signals directly without a scintillator) is clearly preferable but requires high-Z materials to generate sufficient x-ray stopping power in the 20 – 100 keV range. CdTe or CdZnTe CMOS detectors are currently the most attractive options [67], [68] and have been shown to achieve a single-pixel energy resolution of 580 eV at 60 keV [69].

The energy resolution of these semiconductor-based detectors is limited by their semiconducting band gap. Further improved energy resolution would require further reduced bandgaps, in many cases degrading x-ray stopping capabilities. Superconducting detectors, with their significantly smaller band gaps have the clear advantage and can achieve significantly improved energy resolutions.

### *B. Advantages of MKIDs for Energy-Resolved Medical X-Rays*

Both Transition Edge Sensors (TESs) as well as MKIDs (called Thermal Kinetic Inductance Detectors or TKIDs in the x-ray range) have been demonstrated to achieve significantly better energy resolution in the 6 keV range [70], [71]. Up to a $\Delta E$ of 1.8 eV for TES's [70] and 41 eV for TKIDs [72] has been achieved. While significant development efforts have been used to achieve this energy resolution of TES's for soft x-rays, the development of TKIDs is still in a very early stage. As there are no principal limitations [71] that should stop TKIDs from achieving comparable results to the much further developed x-ray TES's, and keeping their superior multiplexibility in mind, TKIDs are the most promising detector technology for energy-resolved medical x-rays.

As superconducting detectors, both TES's and TKIDs come with cryogenic and readout challenges. While TES's typically need to be cooled down to < 80 mK [70], TKIDs operate between 150-200 mK [72] which is easier to achieve with modern cryogenic equipment and are feasible in a clinical environment. Readout complexity on the other hand, is a much larger challenge for TES's compared to TKIDs as TES's require complex and highly sensitive readout electronics at cryogenic temperatures. The use of TKID instruments in hospitals therefore appears less challenging.

Even with the already demonstrated energy resolution of TKIDs, it would be possible to distinguish between a large amount of contrast agents simultaneously in a single-pass x-ray. Both the ionizing irradiation as well as the general diagnostic burden on the patient could therefore be reduced significantly.

In order to achieve good x-ray stopping powers, TKIDs utilize dedicated x-ray absorbers on free floating islands and operate as micro-calorimeters [71]. This allows flexible choice of absorber materials and therefore optimized x-ray absorption for many different x-ray energies. Furthermore, TKIDs have practically no dark or detector noise, further improving imaging contrast.

## V. Synchrotron X-Ray Microscopy

X-ray imaging and spectral analysis is also a powerful tool to study medical and biological cellular samples. In synchrotron light sources, electrons are accelerated to relativistic speeds and distorted with magnetic fields, emitting x-ray radiation typically up to 100 keV [73]. These sources are orders of magnitude brighter compared to typical scientific x-ray sources (e.g. x-ray tubes) [74] and can be highly collimated and tuned to specific wavelengths by fine control of the accelerating magnetic fields in the synchrotron [75]. These x-ray photons can then be focused down to nm scales, allowing for high-resolution mapping of sample structure and, utilizing e.g. characteristic elemental lines, enabling spectroscopic analysis of chemical composition [76].

Medical x-ray imaging at synchrotron facilities has been at the forefront of what is possible for modern x-ray imaging since many decades with more than 50 large scale research facilities worldwide in operation in 2015 [77]. Synchrotron emission has been utilized in many scientific areas including material science, biochemistry and medicine [78], [79]. Synchrotron X-ray Microscopy (SXM) is an important medical diagnostic tool in many cases that require chemical and/or material tracing such as e.g. bone mineralisation mapping, drug uptake in cells or mapping trace metals (Fe, Zn, Cu, …) in brain tissues, implicit in neuro-degenerative diseases (e.g. Parkinson's). SXM is advantageous for soft tissue analyses where contrast is poor in conventional radiography but can be significantly increased with SXM fluorescence, absorption and phase contrast measurements [80]. Typical x-ray energies generated and employed for SXM range from 75 eV up to 25 keV [81], [82].

### *A. State-of-the-art in Synchrotron X-Ray Microscopy (SXM)*

Currently, multiple detector technologies are being employed in synchrotron x-ray microscopy. Indirect detectors (e.g. CCDs or CMOS using scintillators) have been extensively developed for synchrotron microscopy. Their high pixel count allows for high resolution in SXM but they are incapable of achieving good x-ray photon energy resolution without filters and can struggle to reach high frame rates, which is attractive to areas such as fluorescence microscopy or scanning transmission microscopy [83].

Hybrid detectors such as e.g. Medipix [84] are direct conversion detectors that consist of a detecting island (Si, CdTe, …) thick enough to stop incident x-ray photons and separate signal processing circuits. They have small active areas (typically < 70 µm pixel pitch), excellent time resolution (~ 200 ps time stamp) and can be mosaicked to form large arrays, ideal for high-spatial resolution imaging and time-resolved tomography [85], [86], [87]. However, Medipix suffer from poor energy resolution (2.6 keV at 10.5 keV [88]), making them unsuitable for detailed spectroscopic analyses.

Silicon Drift Diodes (SDDs) are also attractive for SXM due to their good energy resolution (144 eV at 5.9 keV), fast sampling rates (< 200 ns) and good noise performance [89], [90]. However, SDDs have large active areas (1 mm$^2$) and thus are limited to single pixel applications, requiring raster scanning. They are also capable of being multiplexed but only to small arrays of up to 1,500 pixels [89].

### *B. Advantages of TKIDs for Synchrotron X-Ray Microscopy*

Superconducting detectors have been developed for synchrotron experiments for more than 10 years and have already achieved significantly improved detector capabilities. Transition Edge Sensors (TESs) (see e.g. [91], [92], [93]) have been successfully integrated into synchrotron beamlines in the past. They operate as micro-calorimeters, exhibit a so-far unmatched energy resolution of up to 1.8 eV at 5.9 keV [70] and offer a very low noise floor. The inherent energy resolution of TESs allows for de-focusing of the synchrotron beam without degrading detector performance, reducing sample radiation exposure and enabling studies of sensitive samples that would otherwise be damaged by the required radiation doses [94]. However, TESs are still constrained to smaller array sizes as they require multi-stage SQUID amplification chains and therefore can't be easily multiplexed.

TKIDs on the other hand offer a unique combination of powerful advantages for SXM: Their pixel size is very flexible and they can utilize a large selection of potential x-ray absorber materials. TKIDs offer single-pixel energy resolution of currently demonstrated 41 eV at 6 keV [72] but are expected to be able to achieve significantly better values with future development efforts [95]. Based on their operating principle, TKIDs are expected to be able to reach energy resolutions comparable to TESs but are significantly easier to multiplex. TKID array sizes in the same range as MKIDs (currently up to 20k pixels [96]) are clearly possible and even megapixel arrays are feasible. TKIDs can achieve better than 1 µs photon arrival time stamping, beneficial in for example, time-resolved tomography. They also offer vanishing noise and a comparably simple pixel design and fabrication procedure. While still in development, TKIDs are one of the most promising detector options for synchrotron x-ray microscopy and will hopefully enable many new scientific insights in the future.

## VI. Conclusion

Microwave Kinetic Inductance Detectors offer significant advantages over existing semiconductor detectors across many bioimaging modalities. These include Fluorescence Microscopy, Near-Infrared Spectroscopy, Energy-Resolved Medical X-Rays and Synchrotron X-ray Microscopy. MKIDs represent a compelling next-generation detector technology for medical based imaging by combining single photon counting, inherent energy resolution and vanishing noise with scalable detectors. With continued future advancements in MKID development, these superconducting detectors have strong potential to improve diagnostic capabilities and patient safety within medical imaging. Despite the challenge of complex cryogenics and readout systems required for MKIDs, similar infrastructures are already feasible in modern clinical environments, making MKIDs a strong candidate for future high-density, superconducting-based imaging systems for low dosage medical diagnostics.

## Acknowledgment

This publication has emanated from research conducted with the financial support of Taighde Éireann – Research Ireland under Grants No. 15/IA/2880 and 21/FFP-P/10213 and also in part by a grant from the Astronomy & Astrophysics Section,

School of Cosmic Physics, Dublin Institute for Advanced Studies.